\documentclass[conference]{IEEEtran}
\usepackage{hyperref}
\usepackage{url}
\IEEEoverridecommandlockouts
\usepackage{eso-pic}   % put in preamble

\usepackage{cite}
\usepackage{amsmath,amssymb,amsfonts}
\usepackage{algorithmic}
\usepackage{graphicx}
\usepackage{textcomp}
\usepackage{xcolor}
\usepackage{booktabs}
\def\BibTeX{{\rm B\kern-.05em{\sc i\kern-.025em b}\kern-.08em
    T\kern-.1667em\lower.7ex\hbox{E}\kern-.125emX}}

\begin{document}

\title{LEAD-Drift: Real-time and Explainable Intent Drift Detection by Learning a Data-Driven Risk Score}

\author{
\IEEEauthorblockN{
Md. Kamrul Hossain\IEEEauthorrefmark{1},
Walid Aljoby\IEEEauthorrefmark{2}
}
\IEEEauthorblockA{
\IEEEauthorrefmark{1}\IEEEauthorrefmark{2}Information and Computer Science Department, King Fahd University of Petroleum and Minerals, Dhahran 31261, Saudi Arabia\\
\IEEEauthorrefmark{2}IRC for Intelligent Secure Systems, King Fahd University of Petroleum and Minerals, Dhahran 31261, Saudi Arabia
}
\IEEEauthorblockA{(e-mail: g202215400@kfupm.edu.sa, waleed.gobi@kfupm.edu.sa)
}
}

% \author{\IEEEauthorblockN{1\textsuperscript{st} Md. Kamrul Hossain}
% \IEEEauthorblockA{\textit{Information and Computer Science Department} \\
% \textit{King Fahd University of Petroleum \& Minerals (KFUPM)}\\
% Dhahran, Saudi Arabia \\
% g202215400@kfupm.edu.sa}
% \and
% \IEEEauthorblockN{2\textsuperscript{nd} Walid Aljoby}
% \IEEEauthorblockA{\textit{Information and Computer Science Department} \\
% \textit{King Fahd University of Petroleum \& Minerals (KFUPM)}\\
% Dhahran, Saudi Arabia \\
% waleed.gobi@kfupm.edu.sa}
% }

\maketitle

\AddToShipoutPictureFG*{%
  \AtPageLowerLeft{%
    \raisebox{40pt}{%
      \makebox[\paperwidth]{%
        \centering\footnotesize
        © 2026 IEEE. Accepted for publication in IEEE ICC 2026.
      }%
    }%
  }%
}

\begin{abstract}
Intent-Based Networking (IBN) simplifies network management, but its reliability is challenged by ``intent drift'', where the network's state gradually deviates from its intended goal, often leading to silent failures. Conventional approaches struggle to detect the subtle, early stages of intent drift, raising alarms only when degradation is significant and failure is imminent, which limits their effectiveness for proactive assurance. To address this, we propose LEAD-Drift, a framework that detects intent drift in real time to enable proactive failure prevention. LEAD-Drift's core contribution is reformulating intent failure detection as a supervised learning problem by training a lightweight neural network on fixed-horizon labels to predict a future risk score. The model's raw output is then smoothed with an Exponential Moving Average (EMA) and passed through a statistically tuned threshold to generate robust, real-time alerts. Furthermore, we enhance the framework with two key features for operational intelligence: a \textbf{multi-horizon modeling} technique for dynamic time-to-failure estimation, and \textbf{per-alert explainability} using SHAP to identify root-cause KPIs. Our evaluation on a time-series dataset shows LEAD-Drift provides significantly earlier warnings, improving the average lead time by 7.3 minutes (+17.8\%) compared to a distance-based baseline. It also reduces alert noise by 80.2\% compared to a weighted-KPI heuristic, with only a minor trade-off in lead time. These results demonstrate that LEAD-Drift as a highly effective, interpretable, and operationally efficient solution for proactive network assurance in IBN.
\end{abstract}


\section{Introduction}
\label{sec:introduction}

Intent-Based Networking (IBN)~\cite{10620891, leivadeas2022survey, rfc9315} is emerging as a transformative paradigm for network automation, promising to align network behavior with high-level business objectives with minimal human intervention. The core functionalities of IBN are twofold\cite{10575429}: \textit{fulfillment}, which translates an intent into network configurations, and \textit{assurance}, which continuously verifies that the network's operational state complies with the intent's target state. While fulfillment has seen significant progress\cite{10622580, 9796918, 10858572}, robust and proactive assurance remains a critical challenge.

\begin{figure}[t!]
    \centering
    % Ensure you have this figure in the specified path
    \includegraphics[width=0.8\columnwidth]{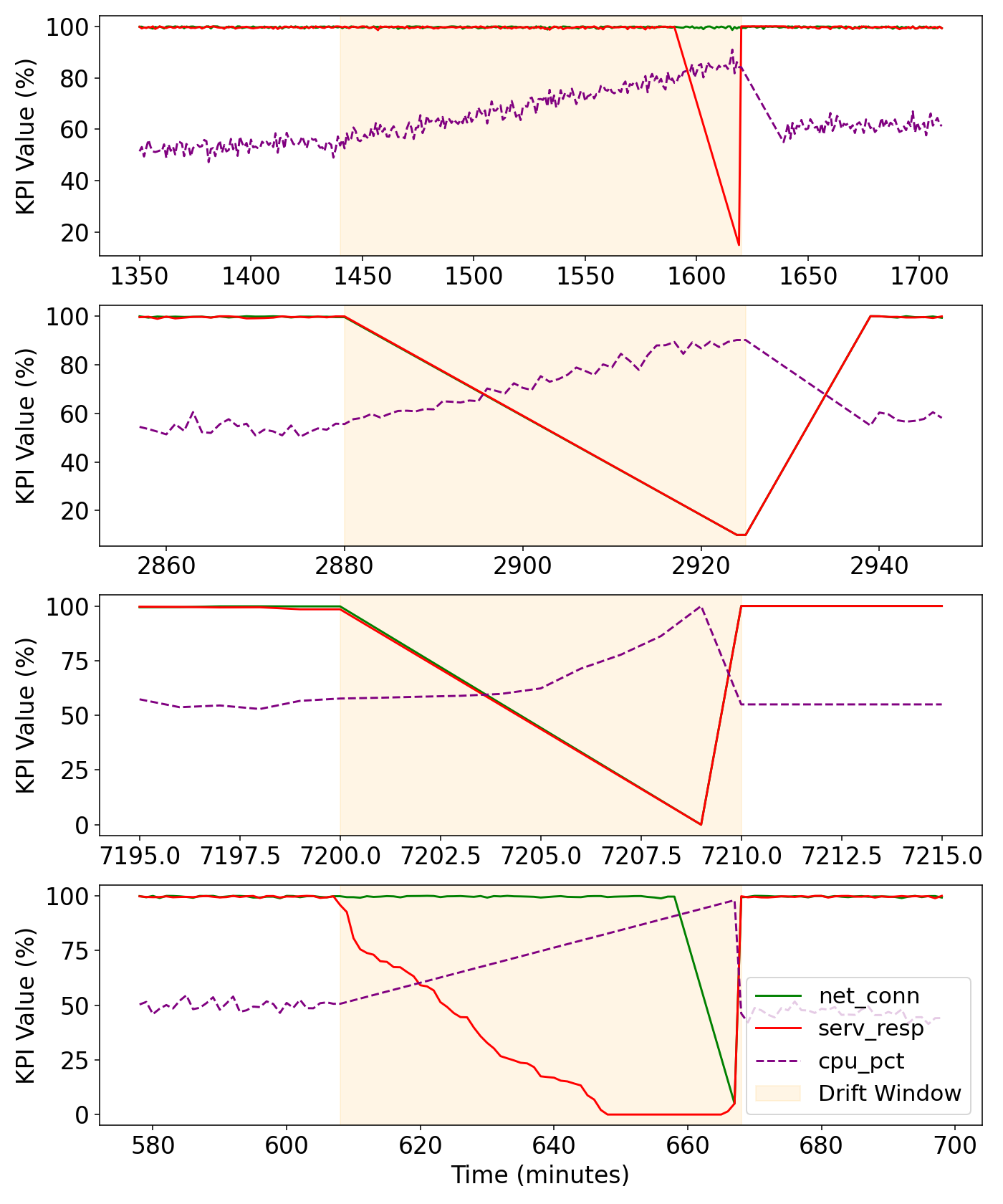}
    \caption{Illustration of a multi-KPI intent drift scenario across four failure cases (top to bottom: resource leak, slow degradation, sudden crash, and imbalance cases). The legend indicates net\_conn (network health), serv\_resp (service responsiveness), and cpu\_pct (CPU utilization).}

    \label{fig:drift_scenarios}
\end{figure}

The primary obstacle to assurance is \textbf{intent drift}\cite{rfc9315}, a gradual and often subtle divergence of the network's operational state from its intended target. This divergence poses a significant threat to network reliability and performance, as shown in Fig.~\ref{fig:drift_scenarios}, which illustrates the evolution of KPIs during drift events of a flow-telemetry collector service that gradually degrades and is eventually restarted. The goal of intent assurance is to detect this drift in real time to prevent performance degradation or outright service failure. For example, proactive drift alerts can prevent quality degradation in multimedia sensing systems~\cite{mowafi2014novel}.

The work in \cite{hossain2025netintent} used iperf and ping tests to verify whether flow rules representing user intents respond as expected. However, this approach is challenging to scale and may interfere with production traffic if not carefully rate-limited. Recent research has begun to address the intent drift problem. Some approaches utilize unsupervised machine learning, such as clustering algorithms like DBSCAN\cite{10770652}, to identify anomalous deviations in network telemetry. Others have proposed mathematical frameworks\cite{10575429} to measure drift as the Euclidean distance between operational and target KPI vectors. While valuable, these approaches have limitations. Unsupervised methods struggle to separate benign anomalies from critical drift, while distance-based methods detect drift only after significant deviation, reducing remediation time.

To overcome these limitations, we introduce \textbf{LEAD-Drift}, a name that reflects our primary goal: achieving a significant \textit{lead time}, i.e., the warning time between an early alert and the failure itself, before an intent failure occurs. We define our framework as learning-enabled early drift-detection, a system that detects drift in real time to enable proactive failure prevention. Instead of detecting a current deviation, LEAD-Drift learns a data-driven, forward-looking risk score. However, a single risk score, while proactive, still presents two operational challenges: it can be opaque to operators, and it provides a static assessment of risk. To solve this, we enhance the framework with two key features for operational intelligence: a \textbf{multi-horizon modeling} technique that provides a dynamic estimate of the time-to-failure, and \textbf{per-alert explainability} using SHAP\cite{10.5555/3295222.3295230} to identify the specific KPIs driving a high-risk prediction.

Our contributions are summarized as follows.
\begin{enumerate}
    \item We reformulate intent failure detection as a supervised, fixed-horizon prediction problem, a novel approach in the context of IBN.
    \item We design and validate an end-to-end, lightweight framework that achieves a 100\% detection rate with an average lead time of 53.6 minutes and a low false positive rate of 0.81 per day in 5-fold cross-validation.
    \item We introduce two features for operational intelligence: per-alert, SHAP-based explainability to make the model's decisions transparent, and a multi-horizon technique to dynamically estimate the time-to-failure.
    \item We demonstrate through comparative analysis that LEAD-Drift offers a superior balance of proactiveness and efficiency over state-of-the-art baseline methods.
\end{enumerate}

\section{Related Work}
\label{sec:related_works}

Intent assurance is a major challenge in realizing IBN, and the specific problem of intent drift which is the gradual, subtle degradation of an intent’s performance remains comparatively underexplored. In particular, reliably detecting drift early enough to prevent service failure is still an emerging research area.

Many intent assurance techniques use machine learning, often via (i) forecasting future KPI values or (ii) classifying near-term network states. For example, Zheng and Leivadeas \cite{9615580} forecast VM CPU usage using LSTMs and trigger alarms when predictions cross fixed thresholds, while Gharbaoui et al. \cite{10942926} use an SVM to classify imminent link congestion from traffic features. Although predictive, these approaches are typically indirect for drift-to-failure prediction, in that they rely on single-KPI forecasts that can be noisy for complex failures or focus on immediate conditions rather than explicitly modeling a future failure event.

Work that directly targets intent drift includes unsupervised telemetry analysis and target-state deviation. Muonagor et al. \cite{10770652} apply clustering (e.g., DBSCAN) to detect emerging anomalous patterns, which is valuable when labeled failures are unavailable. Dzeparoska et al. \cite{10575429} define drift as deviation from a KPI target state and use LLMs to generate corrective policies. However, these mechanisms are not trained to recognize the subtle, multi-variate precursors that predict a failure within a specific future window. Unsupervised clustering may flag any substantial distribution shift, and target-state methods act once KPIs have already deviated.

Our approach builds on fixed-horizon labeling, where models are trained to predict whether a failure will occur within a future time window. This strategy has proven effective for proactive outage prediction in broader network-management settings (e.g., Basikolo et al. \cite{Basikolo2023TowardsZD}). Yet, applying fixed-horizon learning to intent assurance to learn multi-variate precursors of intent failure remains underexplored. LEAD-Drift bridges this gap by formulating intent failure prediction as a supervised, forward-looking learning problem.

\section{The LEAD-Drift Methodology}
\label{sec:methodology}
The LEAD-Drift framework is designed to detect the subtle, incipient patterns of intent drift in real time to enable the proactive prevention of a future intent failure. Our approach is a two-stage process: an offline training phase where a predictive model learns the precursors to failure from historical data, and an online inference phase where the trained model analyzes live network data to generate timely alerts. This section details the core components of our methodology.

\subsection{Problem Formulation}
\label{subsec:problem_formulation}
Let $X_t$ be the vector of network Key Performance Indicators (KPIs) and their derivatives observed at time $t$. An intent failure occurs at a future time $t_f$. The objective is to design a system that raises an alert at time $t_{alert}$ such that the proactive \textbf{lead time}, defined as $T_{lead} = t_f - t_{alert}$, is maximized, while the number of false positive alerts is minimized.

\subsection{Optimization Framework for Proactive Alerting}
\label{subsec:optimization}
Let $s_t$ denote the raw risk score and $S_t$ its EMA-smoothed counterpart (see Eq.~\eqref{eq:ema}). An alert is issued at the \emph{first-crossing} time
\begin{equation}
t_{\text{alert}}(\tau) \;=\; \inf\,\{\, t \,:\, S_t \ge \tau \,\},
\label{eq:talert}
\end{equation}
and, for an episode with failure time $t_f$, the proactive lead time is
\begin{equation}
T_{\text{lead}}(\tau) \;=\; \max\{\, t_f - t_{\text{alert}}(\tau) ,\, 0 \,\}.
\label{eq:leadtime}
\end{equation}

Threshold selection is cast as a Neyman--Pearson style constrained program \cite{Rojo2012} that maximizes expected lead time subject to a false-alarm budget:
\begin{equation}
\begin{aligned}
\max_{\tau \in \mathbb{R}} \quad & \mathbb{E}\!\left[\, T_{\text{lead}}(\tau) \,\right] \\
\text{s.t.} \quad & \Pr\!\left( S_t \ge \tau \,\middle|\, y_t=0 \right) \;\le\; \alpha_{\max},
\end{aligned}
\label{eq:constrained-opt}
\end{equation}
where $y_t$ is the forward-looking label and $\alpha_{\max}\in(0,1)$ is a specified per-step false-alarm cap.

\textbf{Threshold policy:}
Because $S_t$ is monotone in failure likelihood within horizon $H$, the most powerful detector under a fixed per-step false-alarm budget $\alpha_{\max}$ is an upper-threshold rule $\mathbf{1}\{S_t \ge \tau\}$ (Neyman--Pearson). Accordingly, we tune $\tau$ on validation and use it at run time.

%\textbf{Proposition (threshold optimality).} 
%Assume larger $S_t$ implies higher failure risk within the horizon $H$. Among all measurable policies depending on $S_t$ only, there exists a threshold $\tau$ whose upper-level set rule $\mathcal{A}^\star(S_t)=\mathbf{1}\{S_t\ge\tau\}$ attains the maximum in~\eqref{eq:constrained-opt} for any fixed $\alpha_{\max}$. \\
%\emph{Sketch.} Under a fixed false-alarm measure w.r.t.\ $y_t{=}0$, upper-level sets maximize true detections (and thus the lead-time reward) by a Neyman--Pearson argument.\hfill$\square$

\subsection{Forward-Looking Risk Labeling}
\label{subsec:labeling}
The core of our approach is the reformulation of the problem through a forward-looking labeling strategy. This strategy, novel in the context of IBN, transforms historical time-series data into a training set for a predictive model \cite{Basikolo2023TowardsZD, 9557387}. Instead of labeling a data point based on the \textit{current} state of the system, we label it based on a \textit{future} outcome. 

Given a proactive time \textbf{horizon} $H$, we assign a binary label $y_t$ to the feature vector $X_t$ based on whether a failure occurs within the subsequent time window $[t, t+H]$. The label is formally defined as:
\begin{equation}
y_t =
\begin{cases}
1 & \text{if a failure occurs in the interval } [t, t+H] \\
0 & \text{otherwise}
\end{cases}
\label{eq:label}
\end{equation}
This formulation explicitly trains the model to identify patterns in $X_t$ that are precursors to a failure within the next $H$ minutes, making the system inherently proactive.

\subsection{Risk Score Learning}
\label{subsec:model}
The input to our model is a feature vector $X_t$ composed of raw KPIs (e.g., cpu\_pct, ram\_pct, serv\_resp, net\_conn) and engineered features such as the first-order difference ($\Delta$) of key metrics. These delta features capture the rate of change and are crucial for detecting the velocity of drift.

We use a lightweight Multi-Layer Perceptron (MLP) to learn a mapping from the input features to a scalar risk score. The MLP consists of an input layer, two hidden dense layers with ReLU activation functions, and a single linear output neuron. The model is trained to minimize the Mean Squared Error (MSE) loss between its predicted risk score and the target labels $y_t$ derived from Eq.~\eqref{eq:label}. The choice of a simple MLP ensures low computational overhead, making it suitable for real-time inference.

\subsection{Real-Time Detection and Alerting}
\label{subsec:detection}
In the online phase, the trained MLP model receives the live feature vector $X_t$ at each time step and outputs a raw risk score, $s_t$. As this raw score can be noisy, we apply an Exponential Moving Average (EMA) \cite{Hyndman_Athanasopoulos_2021} to produce a smoothed risk score, $S_t$:
\begin{equation}
S_t = \alpha \cdot s_t + (1 - \alpha) \cdot S_{t-1}
\label{eq:ema}
\end{equation}
where $\alpha = 2 / (W+1)$ is the smoothing factor derived from the EMA window size $W$. An alert is triggered if this smoothed risk score $S_t$ exceeds a predetermined alert threshold $\tau$. This two-step process of risk prediction followed by smoothing, makes the alerting system robust against transient noise while remaining sensitive to persistent drift patterns.

\subsection{Alert Threshold Tuning}
\label{subsec:tuning}
The constrained formulation in Eq.~\eqref{eq:constrained-opt} formalizes the objective of maximizing proactive lead time under a false-alarm budget. In practice, an explicit specification of $\alpha_{\max}$ and class-conditional score distributions may be unavailable. A robust and implementation-neutral proxy is therefore adopted: the alert threshold $\tau^\star$ is selected to maximize the \textbf{F1-score} \cite{Manning_Raghavan_Schütze_2008} computed on $\{S_t\}$.

The F1 criterion serves as an empirical surrogate to Eq.~\eqref{eq:constrained-opt}: precision penalizes false positives, recall rewards early detections, and their harmonic mean emphasizes balanced operation. Consistency with deployment is ensured by evaluating decisions with the first-crossing rule in Eq.~\eqref{eq:talert}.

\subsection{Per-Alert Explainability with SHAP}
\label{subsec:shap}
To move beyond black-box predictions and provide actionable insights for network operators, our framework incorporates per-alert explainability. When an alert is triggered, we analyze the specific prediction using \textbf{SHAP (SHapley Additive exPlanations)}\cite{10.5555/3295222.3295230}, a state-of-the-art feature attribution method based on cooperative game theory. For any given prediction, SHAP assigns an attribution value to each input KPI, quantifying its contribution to pushing the model's output away from the baseline. This allows an operator to immediately understand \textit{why} a risk score is high by identifying the specific KPIs that are driving the prediction, making the model's decisions transparent and significantly accelerating the troubleshooting process.

\subsection{Dynamic Lead Time Estimation with Multi-Horizon Models}
\label{subsec:multi_horizon_method}
While a single fixed-horizon model can effectively warn of an impending failure, it provides a static assessment of risk. To offer a more dynamic understanding of an event's urgency, we extend our framework with a \textbf{multi-horizon modeling} technique. This approach involves training a set of specialized LEAD-Drift models, each for a different proactive horizon ($H_1, H_2, \dots, H_n$ where $H_1 < H_2 < \dots < H_n$). In the online phase, by observing which of these models are currently reporting a risk score above their individually tuned thresholds, the system can infer an estimated time-to-failure range. This sequential activation of models provides a dynamic ``countdown" to the failure, giving operators crucial context on how the situation is evolving and how quickly they need to act.

\section{Use Case: Proactive Assurance for a Telemetry Service}
\label{sec:usecase}
We consider a critical network service responsible for ingesting telemetry data, such as NetFlow, with a high-level intent of maintaining \textit{99.99\% operational uptime}. The primary assurance challenge in this scenario is not abrupt failure but gradual performance degradation, or ``brownout'', where subtle, multi-KPI issues like increasing CPU load and degrading service response times precede a full outage. Detecting these complex pre-failure patterns is difficult for traditional assurance methods that rely on simple thresholds for individual KPIs, which motivates the need for a holistic, learning-based system.

To address this, we instantiate the LEAD-Drift framework. The feature vector $X_t$ representing the collector's health is composed of seven KPIs from our simulation environment: CPU, RAM, and storage utilization (\textit{cpu\_pct}, \textit{ram\_pct}, \textit{storage\_pct}); Service Network health (\textit{net\_conn}); Service Response Integrity (\textit{serv\_resp}); and the first-order differences, \textit{cpu\_delta} and \textit{serv\_resp\_delta}, to capture their rate of change. This feature vector is used with the forward-looking labeling strategy (Section~\ref{subsec:labeling}) to train our risk model, which is then deployed for real-time inference as described by our general pipeline in Section~\ref{sec:methodology}.

In an IBN control loop, LEAD-Drift operates in the assurance stage by continuously scoring incoming telemetry and raising early alerts before the uptime intent is violated. Upon an alert, the controller can trigger predefined remediation policies for the telemetry service (e.g., restart/roll back the collector, scale resources, or adjust collection/export parameters) and then continue monitoring to confirm intent re-compliance.

\section{Experimental Evaluation}
\label{sec:evaluation}
We conduct a series of experiments to evaluate the performance of the LEAD-Drift framework. The primary objectives are to quantify its effectiveness in providing early warnings for intent failures and to measure its operational efficiency in terms of false positive rates. We compare our approach against two strong, representative baselines. LEAD-Drift's full implementation is available at \url{https://github.com/Muhammadkamrul/LEAD-Drift}.

\subsection{Dataset Generation}
\label{subsec:dataset_generation}
To facilitate a rigorous and repeatable evaluation, we developed a synthetic dataset generator that models the behavior of a network collector service as described in the example use case in Section~\ref{sec:usecase} at a \textbf{sampling interval of one minute}. This generator produces a controllable time-series dataset designed to emulate common operational characteristics:
\begin{itemize}
    \item \textbf{Diurnal Patterns:} Key KPIs such as CPU and RAM utilization follow sinusoidal patterns to mimic cyclical daily network traffic loads.
    \item \textbf{Benign High-Load Periods:} The simulation injects periods of intense but normal activity (e.g., high CPU and RAM) that are not failures, serving as challenging negative examples to test a model's resilience to false positives.
    \item \textbf{Complex Drift Events:} Failures are modeled as ``imbalance failures'', where a subtle, multi-variate drift precedes a crash. For instance, CPU utilization slowly increases while RAM utilization decreases, and service integrity KPIs become noisy before failing.
\end{itemize}

We generate traces of length $N \in [5{,}000, 100{,}000]$ minutes and export a JSON file containing each injected episode’s drift window $[t_{\text{start}},t_{\text{end}}]$ and failure time $t_f$. Imbalance drift-to-failure episodes are injected approximately once per simulated day ($\lfloor N/1440\rfloor$), and benign high-load periods approximately every two days ($\lfloor N/2880\rfloor$); start times are sampled uniformly at random with a non-overlap guard band.

The model is trained on a vector of seven \textbf{input features}: \textit{cpu\_pct}, \textit{ram\_pct}, \textit{storage\_pct}, \textit{net\_conn}, \textit{serv\_resp}, and two engineered features, \textit{cpu\_delta} and \textit{serv\_resp\_delta}. The delta features represent the first-order difference of their respective KPIs and provide the model with crucial information about their rate of change. The \textbf{target feature} is a proactive, binary label derived using the fixed-horizon strategy detailed in Section~\ref{subsec:labeling}, where a label of `1' indicates that a failure is imminent within the defined future horizon.

\subsection{Limitations and Generalizability}
\label{subsec:limitations}
While the synthetic generator enables repeatable drift injection with precise ground truth, it cannot fully capture the heterogeneity of operational networks (e.g., missing/noisy telemetry, overlapping incidents, and previously unseen failure modes). Therefore, the reported lead time and false-positive rates should be interpreted as performance under controlled conditions. Nevertheless, LEAD-Drift is intent-agnostic; the same fixed-horizon labeling and lightweight risk scoring pipeline can be trained on other KPI sets available from standard monitoring stacks, provided failure/incident annotations (or proxy failure signals) exist.

\subsection{Implementation and Training}
\label{subsec:training}
\begin{figure}[t!]
    \centering
    \includegraphics[width=\columnwidth]{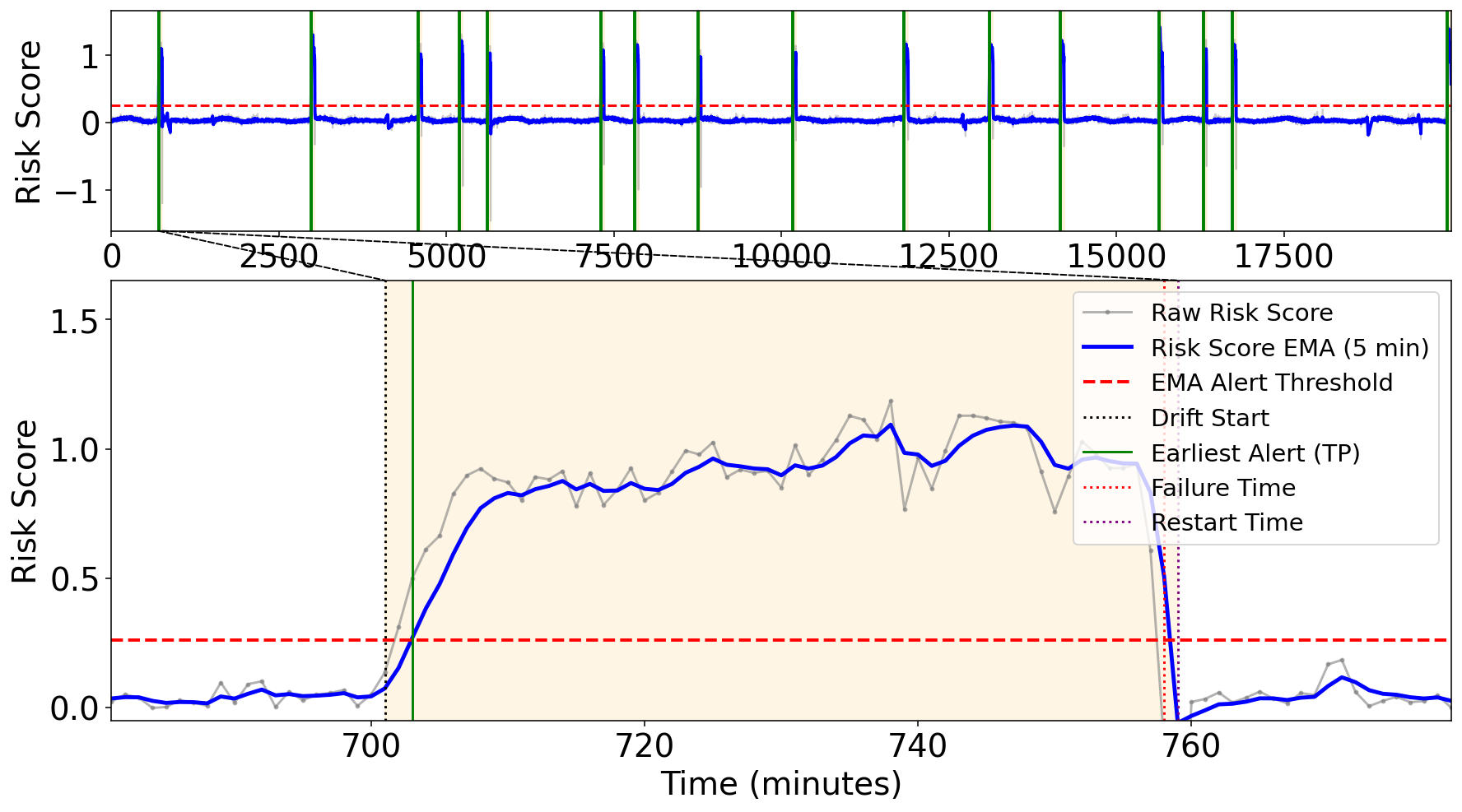}
    \caption{Example intent drift detection timeline. The upper panel is a bird’s-eye view of the whole timeline, while the zoomed panel shows detailed behavior around one event.}
    \label{fig:example_detection_timeline}
\end{figure}

As described in Section~\ref{sec:methodology}, our risk model is a lightweight MLP. The model was trained with a batch size of 512 for 25 epochs. A key step in our data preparation pipeline was to split the data into training and validation sets \textit{before} calculating the delta features, thereby preventing any data leakage from the test set into the training process. The raw risk scores produced by the model during inference were smoothed using an EMA with a window size ($W$) of 5, corresponding to a \textbf{5-minute time window (5 samples)}. The experiments were swept across dataset sizes ranging from 5,000 to 100,000 minutes to evaluate performance under varying data availability. In Fig. \ref{fig:example_detection_timeline}, a snapshot from a random fold of intent drift detection timeline is illustrated.

We report 5-fold cross-validation results with an alert threshold $\tau$ tuned per fold on the training split by maximizing F1-score on the EMA-smoothed scores. The tuned $\tau$ is then applied to the held-out fold using the first-crossing alert rule. False positives/day is computed as the total number of alerts outside annotated drift windows divided by the evaluated duration (in days).

\subsection{Performance Analysis}
\label{subsec:results}

\subsubsection{Cross validation outcome}

\begin{table}[t!]
\centering
\caption{Model Performance in 5-Fold Cross-Validation}
\label{tab:model_performance}
\footnotesize
\begin{tabular}{@{}lc@{}}
\toprule
\textbf{Metric} & \textbf{Value} \\ \midrule
Detection Rate & 100.00\% \\
Average Lead Time (min) & 53.62 $\pm$ 6.28 \\
False Positive Rate/day & 0.81 \\ \bottomrule
\end{tabular}
\end{table}

The 5-fold cross-validation results, summarized in Table~\ref{tab:model_performance}, confirm that the LEAD-Drift framework is highly effective, proactive, and efficient. The model achieved a perfect 100\% Detection Rate, successfully identifying every failure event across all test folds. Critically, the framework demonstrated a strong proactive capability, providing an average lead time of 53.62 minutes before failure. The low standard deviation of 6.28 minutes indicates this performance is stable and consistent across different data subsets. This high level of proactiveness was achieved with excellent operational efficiency, as shown by the low false positive rate of just 0.81 alarms per day, confirming the model's ability to balance early detection with high precision.

\subsubsection{Explainability}

\begin{figure}[t!]
    \centering
    \includegraphics[width=0.8\columnwidth]{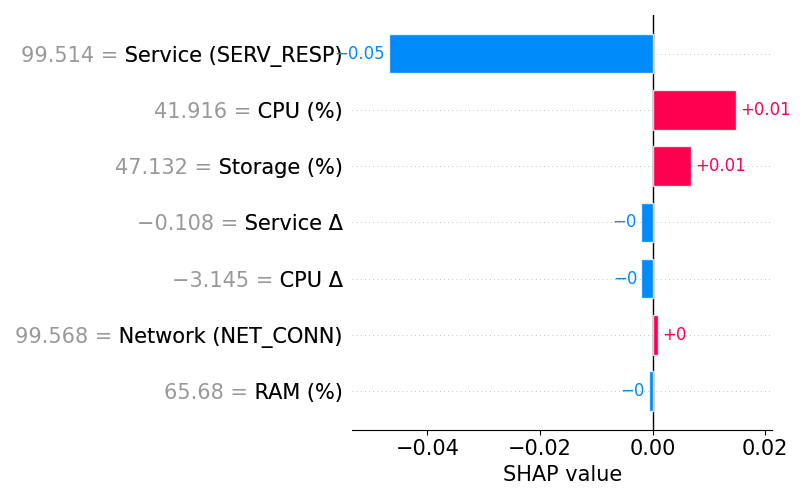}
    \caption{SHAP analysis of a healthy test instance. The serv\_resp KPI provides a strong negative contribution, correctly driving the risk score down despite minor positive contributions from other KPIs.}
    \label{fig:shap_force_plot}
\end{figure}

To demonstrate local explainability, we applied SHAP to a healthy test instance (index 95,000), for which the model correctly predicted a low risk score of 0.0182. The analysis in Fig. \ref{fig:shap_force_plot} reveals the model's reasoning: the strong negative (safe) contribution from the Service Response Integrity (serv\_resp) KPI (-0.0468) overwhelmingly counteracted minor positive (risky) contributions from cpu\_pct (+0.0148) and storage\_pct. This confirms the model's ability to holistically weigh KPIs and make an interpretable decision.

%To demonstrate the local explainability of the LEAD-Drift framework, we performed a feature attribution analysis using SHAP on a representative data instance (index 95,000) from the test set. For this instance, the model predicted a very low risk score of 0.0182, correctly identifying a healthy system state.
%The SHAP analysis, which is visualized in a force plot in Fig. \ref{fig:shap_force_plot}, reveals the model's reasoning behind this prediction. The dominant feature influencing the outcome was the Service Response Integrity (serv\_resp), which contributed a strong negative SHAP value of -0.0468, significantly pushing the prediction towards a safe state. This strong signal of health effectively counteracted the minor, positive (risk-increasing) contributions from elevated CPU utilization (cpu\_pct: +0.0148) and storage usage (storage\_pct: +0.0069).
%This example demonstrates the model's ability to weigh multiple KPIs holistically. It correctly learned to prioritize the excellent health of a critical indicator like serv\_resp over minor fluctuations in other metrics, resulting in a robust and interpretable decision.

\subsubsection{Dynamic Lead Time Estimation with Multi-Horizon Models}
\begin{figure}[t!]
    \centering
    \includegraphics[width=\columnwidth]{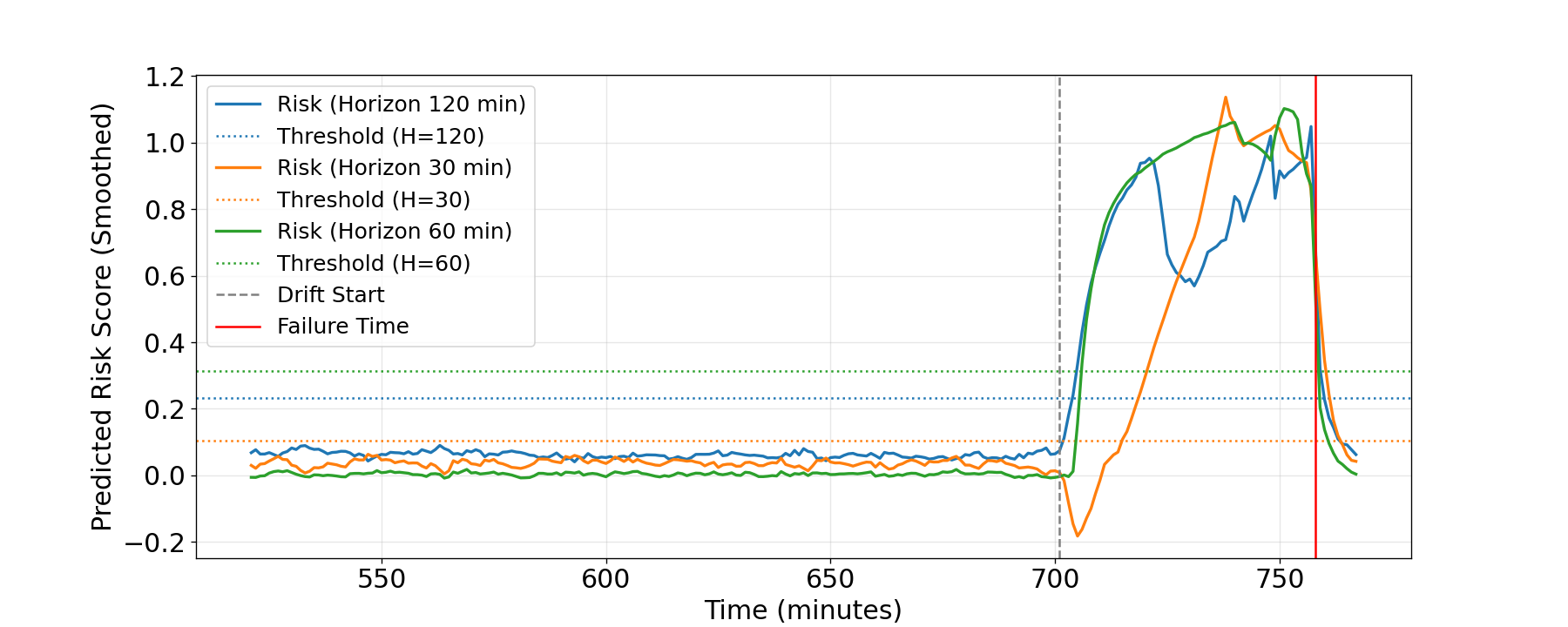}
    \caption{Multi-horizon risk scores for a drift event. The models for H=120, H=60, and H=30 minutes sequentially cross their individually tuned thresholds (dotted lines) as the failure time (red line) approaches.}
    \label{fig:multi_horizon}
\end{figure}

We trained three models with horizons $H\in\{120,60,30\}$ minutes, each with its own tuned threshold. As shown in Fig.\ref{fig:multi_horizon}, the models alert sequentially as failure approaches, yielding a tightening TTF range (e.g., alert at $H{=}120$ but not $H{=}60$ implies $\widehat{\mathrm{TTF}}\in(60,120]$ minutes). This provides a practical ``countdown" view of urgency beyond a single binary alert.

\subsection{Comparison with Baselines}
\label{subsec:comparison}

We compare LEAD-Drift against two strong baselines. The first, which we call ``Weighted KPIs'', is a sophisticated heuristic that represents a common research pattern of using importance scores from a primary model to guide a simpler predictor. This importance-guided approach is demonstrated in related work for both network SLA analysis by Terra et al.\cite{9322496} and for large-scale hardware failure prediction by Han et al.\cite{9505157}. The second baseline\cite{10575429} implements the mathematical distance-based drift detection method. %Table~\ref{tab:results} summarizes the average performance improvements of LEAD-Drift over these Baselines across different dataset sizes.

% \begin{table}[t!]
% \centering
% \caption{Average Performance Improvement of LEAD-Drift}
% \label{tab:results}
% \resizebox{\columnwidth}{!}{%
% \begin{tabular}{@{}lcc@{}}
% \toprule
% \textbf{Comparison Baseline} & \textbf{Lead Time Improv. (min)} & \textbf{FP Reduction (\#/day)} \\ \midrule
% Weighted-KPI heuristic Baseline   & -3.38                      & +4.12                 \\
% Distance-based Baseline  & +7.30                     & +0.29                \\ \bottomrule
% \end{tabular}%
% }
% \end{table}

\subsubsection{Analysis vs. Weighted-KPI heuristic Baseline}
The comparison with the weighted-KPI heuristic baseline reveals a crucial trade-off. LEAD-Drift demonstrates its superior performance by achieving a \textbf{80.2\% reduction in false positives}, cutting the false alarm rate by an average of 4.12 alerts per day. It achieves this remarkable improvement in operational efficiency with only a minor and acceptable compromise in proactiveness (a lead time reduction of 3.38 minutes). This shows that LEAD-Drift is not simply reacting to large KPI swings; it has learned to distinguish the true, subtle signatures of an impending failure from benign, noisy fluctuations that can fool a less sophisticated heuristic.

\subsubsection{Analysis vs. Distance-based Baseline}
The comparison with the distance-based baseline shows that LEAD-Drift offers a substantial advantage in proactiveness. By learning the predictive patterns of pre-failure states, it provides an alert an average of \textbf{7.3 minutes earlier} than the distance-based method. This represents a 17.8\% increase in lead time, which is a critical gain for any proactive assurance system, allowing operators significantly more time for remediation. This highlights the inherent advantage of a learned, forward-looking risk model over a method that detects deviations from a static, predefined `healthy' state.

\subsubsection{Implications}

\begin{figure}[t!]
    \centering
    \includegraphics[width=\columnwidth]{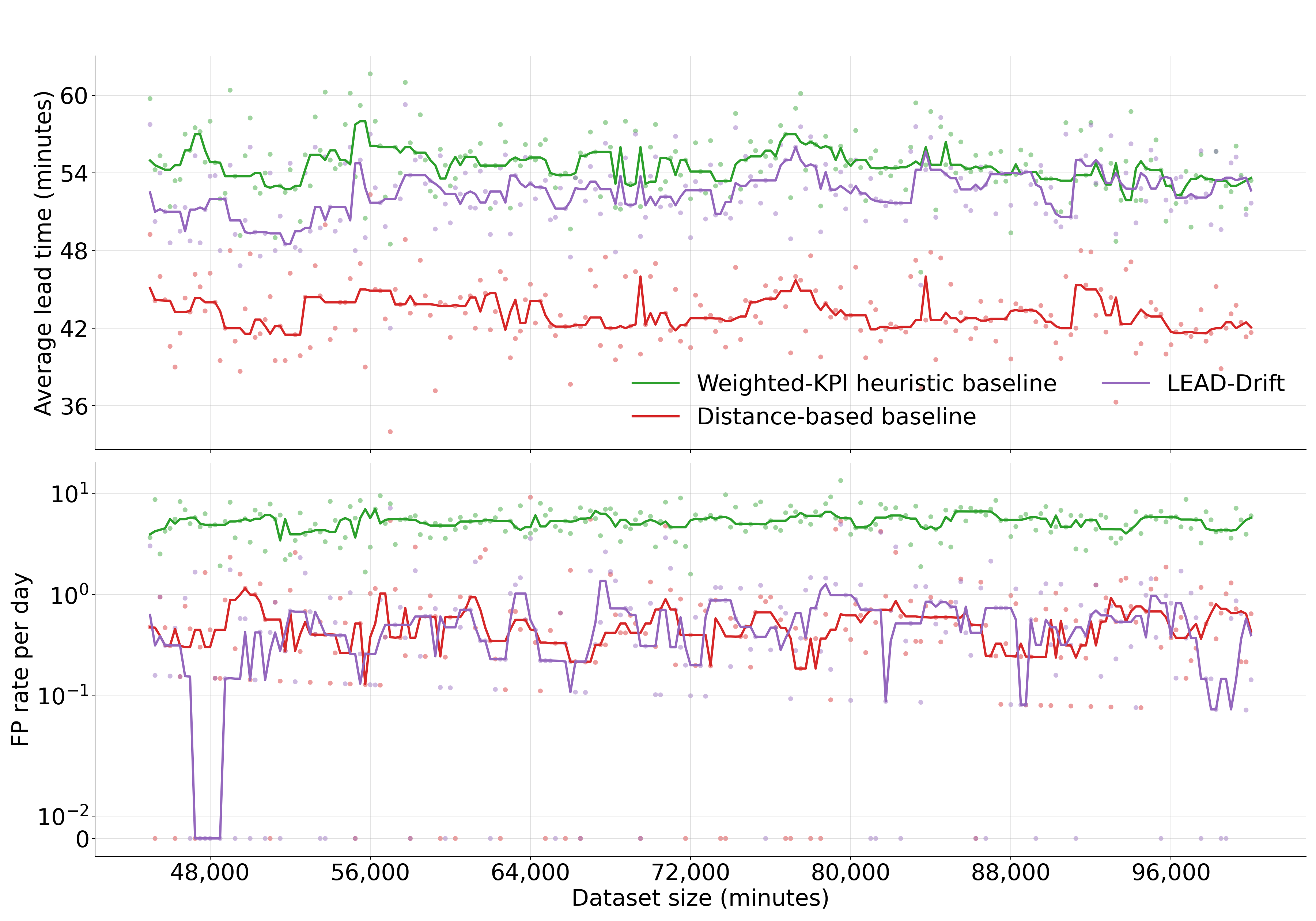}
    \caption{Performance of LEAD-Drift vs. baselines across varying dataset sizes, showing stable lead time and false positive rates.}
    \label{fig:comparison}
\end{figure}

\begin{figure}[t!]
    \centering
    \includegraphics[width=\columnwidth]{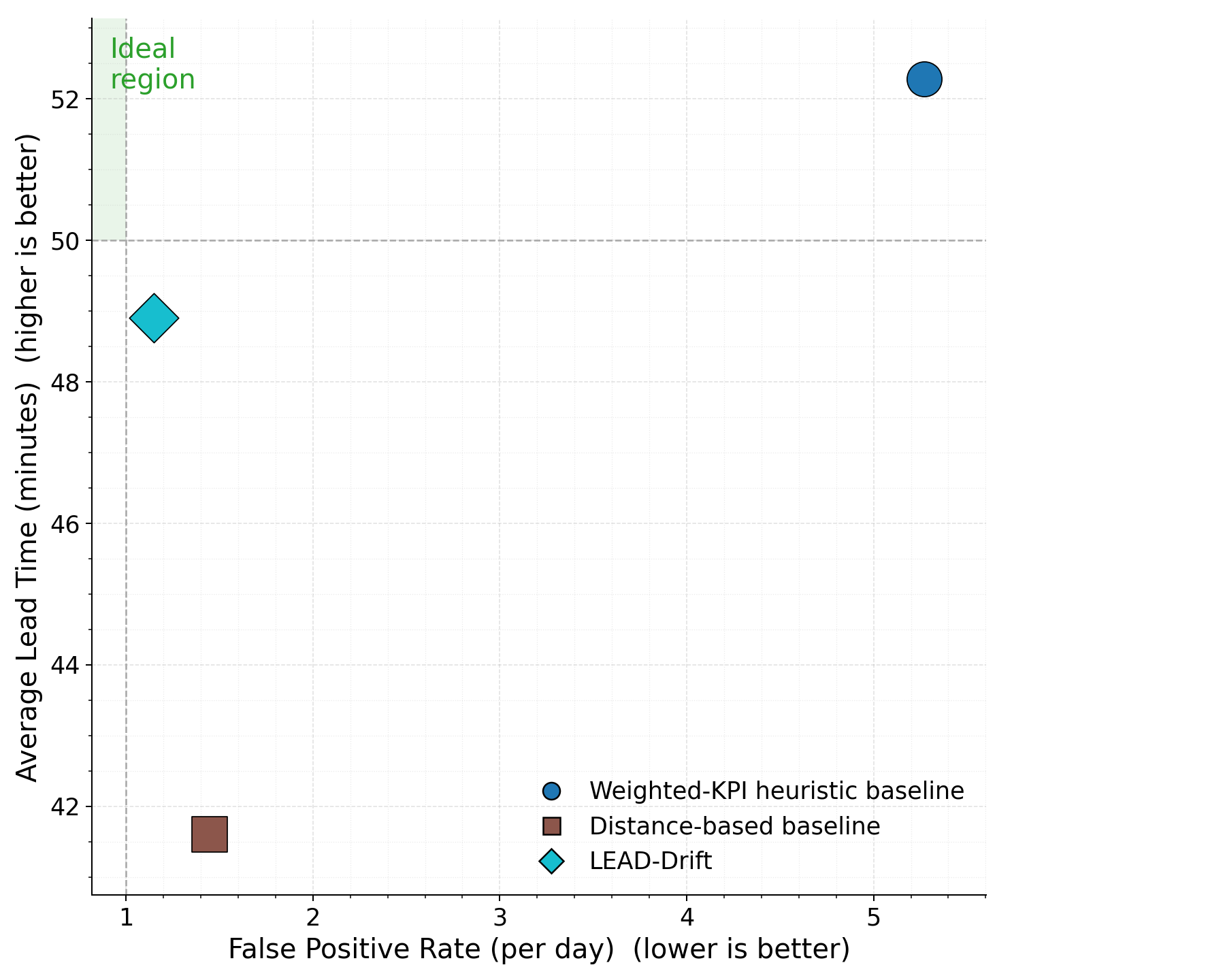}
    \caption{Performance trade-off analysis. LEAD-Drift operates closest to the ideal region (top-left), demonstrating the best balance of high lead time and low false positive rate.}
    \label{fig:tradeoff}
\end{figure}

In Fig. \ref{fig:comparison}, we illustrated the average lead time and false positive rate per day of the evaluated methods across different dataset sizes. In Fig. \ref{fig:tradeoff}, the tradeoff between lead time and false positive is illustrated which shows that LEAD-Drift is the nearest to the ideal region among all methods. This evaluation clearly demonstrates that LEAD-Drift provides the best overall balance of proactiveness and operational efficiency. It delivers significantly earlier warnings than statistical methods while being far more precise and reliable than advanced heuristics, making it a highly effective and practical solution for intent assurance.

\section{Conclusion}
\label{sec:conclusion}
In this paper, we addressed the critical challenge of proactive intent assurance in IBN by introducing \textbf{LEAD-Drift}, a novel framework that reformulates intent failure prediction as a supervised, forward-looking learning problem. Our extensive evaluation demonstrated that LEAD-Drift provides a superior balance of proactiveness and efficiency, significantly improving detection lead time while reducing false positives compared to state-of-the-art baselines. Furthermore, we enhanced the framework with SHAP-based explainability for transparent, actionable alerts and a multi-horizon modeling technique to provide a dynamic ``countdown" to failure. This work represents a significant step towards more autonomous, reliable, and interpretable network assurance systems. Future work will focus on extending this framework to handle multiple, conflicting intent types and validating its performance in a real-world IBN controller.

\section*{Acknowledgment}
We would like to acknowledge the support provided by the Deanship of Research and IRC-ISS at King Fahd University of Petroleum \& Minerals (KFUPM).

\bibliographystyle{IEEEtran}
\bibliography{references}

@INPROCEEDINGS{10770652,
  author={Muonagor, Chukwuemeka and Bensalem, Mounir and Jukan, Admela},
  booktitle={Proc. IEEE LATINCOM}, 
  title={Performance Analysis of Learning-based Intent Drift Detection Algorithms in Next Generation Networks}, 
  year={2024},
  volume={},
  number={},
  pages={1-6},
  doi={10.1109/LATINCOM62985.2024.10770652}}

@INPROCEEDINGS{10942926,
  author={Gharbaoui, M. and Martini, B. and Berardi, D. and Castoldi, P.},
  booktitle={Proc. IEEE ICIN}, 
  title={Towards Intent Assurance: A Traffic Prediction Model for Software-Defined Networks}, 
  year={2025},
  volume={},
  number={},
  pages={135-139},
  doi={10.1109/ICIN64016.2025.10942926}}

@INPROCEEDINGS{9615580,
  author={Zheng, Xiaoang and Leivadeas, Aris},
  booktitle={Proc. IEEE CNSM)}, 
  title={Network Assurance in Intent-Based Networking Data Centers with Machine Learning Techniques}, 
  year={2021},
  volume={},
  number={},
  pages={14-20},
  doi={10.23919/CNSM52442.2021.9615580}}

@book{Hyndman_Athanasopoulos_2021, place={Melbourne}, edition={3rd}, title={Forecasting: Principles and practice}, publisher={OTexts}, author={Hyndman, Rob J. and Athanasopoulos, George}, year={2021}}

@inproceedings{10.5555/3295222.3295230,
author = {Lundberg, Scott M. and Lee, Su-In},
title = {A unified approach to interpreting model predictions},
year = {2017},
isbn = {9781510860964},
publisher = {Curran Associates Inc.},
address = {Red Hook, NY, USA},
booktitle = {Proc. NIPS},
pages = {4768–4777},
numpages = {10},
location = {Long Beach, California, USA},
series = {NIPS'17}
}

@book{Manning_Raghavan_Schütze_2008, place={Cambridge}, title={Introduction to Information Retrieval}, publisher={Cambridge University Press}, author={Manning, Christopher D. and Raghavan, Prabhakar and Schütze, Hinrich}, year={2008}}

@INPROCEEDINGS{9557387,
  author={Züfle, Marwin and Agne, Joachim and Grohmann, Johannes and Dörtoluk, Ibrahim and Kounev, Samuel},
  booktitle={Proc. IEEE INDIN}, 
  title={A Predictive Maintenance Methodology: Predicting the Time-to-Failure of Machines in Industry 4.0}, 
  year={2021},
  volume={},
  number={},
  pages={1-8},
  doi={10.1109/INDIN45523.2021.9557387}}

@Inbook{Rojo2012,
author="Rojo, Javier",
editor="Rojo, Javier",
title="On Testing of Hypotheses",
bookTitle="Selected Works of E. L. Lehmann",
year="2012",
publisher="Springer US",
address="Boston, MA",
pages="89--94",
isbn="978-1-4614-1412-4",
doi="10.1007/978-1-4614-1412-4_11"
}

@article{hossain2025netintent,
  title={NetIntent: Leveraging Large Language Models for End-to-End Intent-Based SDN Automation},
  author={Hossain, Md Kamrul and Aljoby, Walid},
  journal={IEEE Open Journal of the Communications Society},
  volume={6},
  pages={10512--10541},
  year={2025},
  publisher={IEEE}
}

@article{Basikolo2023TowardsZD,
  title={Towards zero downtime: Using machine learning to predict network failure in 5G and beyond},
  author={Emmanuel Basikolo and Thomas Basikolo},
  journal={ITU Journal on Future and Evolving Technologies},
  volume={4},
  number={3},
  pages={434--446},
  year={2023},
  publisher={ITU},
  doi = "10.52953/PYAF8065"
}

@misc{rfc9315,
    series =    {Request for Comments},
    number =    9315,
    howpublished =  {RFC 9315},
    publisher = {RFC Editor},
    doi =       {10.17487/RFC9315},
    url =       {https://www.rfc-editor.org/info/rfc9315},
    author =    {Alexander Clemm and Laurent Ciavaglia and Lisandro Zambenedetti Granville and Jeff Tantsura},
    title =     {{Intent-Based Networking - Concepts and Definitions}},
    pagetotal = 23,
    year =      2022,
    month =     oct
}

@INPROCEEDINGS{10622580,
  author={Mostafa, Salwa and Elbamby, Mohammed S. and Abdel-Aziz, Mohamed K. and Bennis, Mehdi},
  booktitle={Proc. IEEE ICC}, 
  title={Intent Profiling and Translation Through Emergent Communication}, 
  year={2024},
  volume={},
  number={},
  pages={5281-5286},
  doi={10.1109/ICC51166.2024.10622580}}

@INPROCEEDINGS{9505157,
  author={Xu, Fan and Han, Shujie and Lee, Patrick P. C. and Liu, Yi and He, Cheng and Liu, Jiongzhou},
  booktitle={Proc. IEEE/IFIP DSN}, 
  title={General Feature Selection for Failure Prediction in Large-scale SSD Deployment}, 
  year={2021},
  volume={},
  number={},
  pages={263-270},
  doi={10.1109/DSN48987.2021.00039}}

@INPROCEEDINGS{9322496,
  author={Terra, Ahmad and Inam, Rafia and Baskaran, Sandhya and Batista, Pedro and Burdick, Ian and Fersman, Elena},
  booktitle={Proc. IEEE GLOBECOM}, 
  title={Explainability Methods for Identifying Root-Cause of SLA Violation Prediction in 5G Network}, 
  year={2020},
  volume={},
  number={},
  pages={1-7},
  doi={10.1109/GLOBECOM42002.2020.9322496}}

@article{leivadeas2022survey,
  title={A survey on intent-based networking},
  author={Leivadeas, Aris and Falkner, Matthias},
  journal={IEEE Communications Surveys \& Tutorials},
  volume={25},
  number={1},
  pages={625--655},
  year={2022},
  publisher={IEEE}
}

@INPROCEEDINGS{10575429,
  author={Dzeparoska, Kristina and Tizghadam, Ali and Leon-Garcia, Alberto},
  booktitle={Proc. IEEE NOMS}, 
  title={Intent Assurance using LLMs guided by Intent Drift}, 
  year={2024},
  volume={},
  number={},
  pages={1-7},
  doi={10.1109/NOMS59830.2024.10575429}}

@INPROCEEDINGS{10858572,
  author={Zheng, Naigong and Li, Fuliang and Li, Ziming and Yang, Yu and Hao, Yimo and Liu, Chenyang and Wang, Xingwei},
  booktitle={Proc. IEEE ICNP}, 
  title={Configtrans: Network Configuration Translation Based on Large Language Models and Constraint Solving}, 
  year={2024},
  volume={},
  number={},
  pages={1-12},
  doi={10.1109/ICNP61940.2024.10858572}}

@INPROCEEDINGS{10620891,
  author={Ouyang, Ying and Li, Changle and Zhang, Jingwen and Zhao, Xiaoxue and Yang, Chungang},
  booktitle={Proc. IEEE INFOCOM Workshop}, 
  title={Intent-Driven 6G End-to-End Network Orchestration}, 
  year={2024},
  volume={},
  number={},
  pages={1-2},
  doi={10.1109/INFOCOMWKSHPS61880.2024.10620891}}

@INPROCEEDINGS{9796918,
  author={Collet, Alan and Banchs, Albert and Fiore, Marco},
  booktitle={Proc. IEEE INFOCOM}, 
  title={LossLeaP: Learning to Predict for Intent-Based Networking}, 
  year={2022},
  volume={},
  number={},
  pages={2138-2147},
  doi={10.1109/INFOCOM48880.2022.9796918}}

@article{mowafi2014novel,
  title={A novel approach for extracting spatial correlation of visual information in heterogeneous wireless multimedia sensor networks},
  author={Mowafi, Moad Y and Awad, Fahed H and Aljoby, Walid A},
  journal={Computer Networks},
  volume={71},
  pages={31--47},
  year={2014},
  publisher={Elsevier}
}

\end{document}